\documentstyle[12pt]{article}
\renewcommand{\title}[1]{\large\bf \mbox{}\\ \mbox{}\\ \mbox{}\\ \mbox{}\\
     #1\bigskip\medskip\\}
\renewcommand{\author}[1]{\large #1\\ \smallskip}

\newcommand{\hs}[1]{\mbox{\hspace*{#1cm}}}

\newcommand{\address}[1]{{\narrower\normalsize\it #1\\}\bigskip}
\newcommand{\Z}{\mbox{\sf Z\hspace*{-0.45em}Z}}

\def\wt#1#2#3#4#5#6{#1(#2,#3,#4,#5|#6)}

\def\ZZ{{\Z}}
\def\scr{\scriptsize}

\def\and{\;\;{\rm and}\;\;}

\def\re{\mbox{${\Re}e$ }}
\def\im{\mbox{${\Im}m$ }}
\def\i{\mbox{\small i}}
\def\-{\!-\!}
\def\+{\!+\!}
\def\<{\langle}
\def\>{\rangle}
\def\({\biggl(}
\def\){\biggr)}
\def\h{\hspace*{0.5cm}}

\def\half {\mbox{$\textstyle {1 \over 2}$}}
\def\talf {\mbox{$\textstyle {3 \over 2}$}}
\def\mat {\pmatrix}
\def\smat#1{\mbox{\small $\mat{#1}$}}
\def\ba{\begin{array}}
\def\ea{\end{array}}
\def\be{\begin{eqnarray}}
\def\ee{\end{eqnarray}}

\def\ol{\overline}
\def\no{\nonumber}
\def\disp{\displaystyle}

\def\T{\mbox{\boldmath { $ T$}}}

\def\1{\mbox{\boldmath I}}

\def\a{\mbox{$\alpha$}}
\def\U{\mbox{$\cal A$}}

\def\la{\mbox{${l}\!a$}}

\def\lA{\mbox{${l}\!A$}}
\def\[{\mbox{\huge [}}
\def\]{\mbox{\huge ]}}
\def\lam{\lambda}
\def\lp#1{\mbox{$\ell #1$}}
\def\llp#1{\mbox{$\ell\lam #1$}}
\def\reg{\mbox{\scr
   I\hspace*{-0.05cm}I\hspace*{-0.05cm}I/I\hspace*{-0.05cm}V }}
\begin{document}

\newlength{\origbaselineskip}
\setlength{\origbaselineskip}{\baselineskip}

\begin{flushright}
27/01/95\\
\today
\end{flushright}
\begin{center}
\title{Further solutions of  critical  ABF RSOS models}
\author{Yu-kui Zhou\footnote{Email:
        \mbox{\sl zhouy@maths.anu.edu.au}}$^,$\footnote{
      On leave of absence from {\sl Institute of Modern Physics,
                       Northwest University, Xian 710069, China}}}
\address{Mathematics Department, University of Melbourne,
         \\ Parkville, Victoria 3052, Australia \\ and \\
Department of Mathematics, The Australian National University
\\ Canberra, ACT 0200, Australia\footnotemark}
\footnotetext{Postal address.}

\begin{abstract}
The restricted SOS model of Andrews, Baxter and Forrester has been
studied.  The finite size corrections to the eigenvalue spectra of
the transfer matrix of the model with a more general crossing
parameter have been calculated. Therefore the conformal weights and
the central charges of the non-unitary or unitary minimal conformal
field have been extracted from the finite size corrections.

\end{abstract}
\end{center}

\bigskip\bigskip
hep-th/9504107

revised version

\section{Introduction}\setcounter{equation}{0}

The ABF restricted solid-on-solid (RSOS) model has been found by
Andrews, Baxter and Forrester (ABF)
in  1984 \cite{ABF:84}. It has been well known that the model provides
realizations of  unitary minimal conformal field theories
\cite{BPZ:84,Huse:84,FQS:84}. This has been further confirmed
by studying the finite-size corrections
to the ground state energy  \cite{Affleck:86}--\cite{KlPe:92}
%\cite{Affleck:86,KiRe:87,BaRe:89,KlPe:91,KuSu:92,KlPe:92}
(also  see \cite{DeWo:85}--\cite{Zhou:95}
%\cite{DeWo:85,Wo:87,GeRi:87,DeKa:87,HQB:87,Ka:88,
%ABB:87,CIZ:87,KBP:91,ZhPe:95,Zhou:95}
for related works). Among these works,  much effort has been focused on
the ABF model corresponding to the unitary minimal conformal field
theories. By contrast, the finite-size corrections to the transfer
matrix of the ABF model corresponding to  non-unitary minimal
conformal field theories  have received no attention.

The local height probabilities of the ABF model with a crossing
parameter $\lam=k\pi/h$, where two relatively prime positive integers
satisfy $k<h$, have been calculated in \cite{FoBa:85}. In this paper,
with the same motivation, we repeat the consideration of the
finite-size correction calculation of the ABF model with $\lam=k\pi/h$.
In general the model will no longer be physical as there will be some
negative face weights. Nevertheless, the non-unitary minimal conformal
field theories \cite{FQS:84} can be realized as the critical
continuum of the ABF RSOS model with the crossing parameter
$\lam=k\pi/h$. The model is therefore of independent interest for
this feature.

In \cite{KBP:91} an analytic method has been presented to find the
finite-size corrections involving the central charges for the
six-vertex model with a twisted boundary condition. The method has
been successfully applied to the other models (see \cite{WBN:92}
for example). In these works only the central charges have been obtained.
In fact the central charge and conformal weights together could
appear in the finite-size corrections to the eigenvalue spectra of the
transfer matrix. In this paper, following the calculation
presented in \cite{KBP:91}, we find the finite-size corrections to the
eigenvalue spectra of transfer matrices of the critical ABF RSOS model.
{}From the corrections both the central charges and the conformal weights
of non-unitary minimal conformal field theories are extracted. This
generalizes the method presented in \cite{KBP:91} to find the conformal
weights of the ABF SOS model.

We first review the ABF RSOS model and
the Bethe ansatz solutions of transfer matrices
briefly in section~\ref{model}. In
section~\ref{calculation} we find an integral
nonlinear equation and express the finite-size corrections
in terms of the solution of the nonlinear equation. Then the
effective central charges including the conformal weights
are extracted from the finite-size corrections.
A brief discussion is presented in the final section.

\subsection{Models and Bethe ansatz solutions}\label{model}
\setcounter{equation}{0}
The ABF RSOS model can be given by Baxter's SOS model, which was
introduced in order to solve the eight-vertex model with the $R$-matrix
\be
R(u)=\smat{a(u)&0&0&d(u)\cr
           0&b(u)&c(u)&0\cr
           0&c(u)&b(u)&0\cr
           d(u)&0&0&a(u)} \;
\ee
where
\be
a(u)=\Theta(\lam)\Theta(u)H(u+\lam)\;, &&
b(u)=\Theta(\lam)H(u)\Theta(u+\lam)\;,  \no\\
c(u)=H(\lam)\Theta(u)\Theta(u+\lam)\;,   &&
d(u)=H(\lam)H(u)H(u+\lam) \; .
\ee
The $R$-matrix satisfies the Yang-Baxter equation
\cite{Yang:67,Baxter:72}
\be
R^{12}(u)R^{13}(u+v)R^{23}(v)&=&R^{23}(v)R^{13}(u+v)R^{12}(u)\; .
\label{YBE}
\ee
Baxter has shown in \cite{Baxter:72} that the eight-vertex model
%the transfer matrix of the eight-vertex model
%\be
%\T_{8-v}(u)={\rm tr}_c\; R^{c,1}(u-\half\lam)R^{c,2}(u-\half\lam)
%       \cdots,R^{c,N}(u-\half\lam)
%\ee
%can be transferred into the row-to-row transfer matrix of
can be transferred into the SOS model, which is defined by
 the following face weights
\be
W(\lp{\pm 1},\lp{\pm 2},\lp{\pm 1},\ell|u)
         &=&{h(u+\lam)\over h(\lam)} \no\\
W(\lp{\mp 1},\ell,\lp{\pm 1},\ell|u)
         &=&{h(\xi+\llp{\pm \lam})h(u)\over h(\xi+\ell\lam)h(\lam)}
      \label{fw} \\
W(\lp{\pm 1},\ell,\lp{\pm 1},\ell|u)
         &=&{h(\xi+\llp{\mp u})\over h(\xi+\ell\lam)} \no
\ee
where the height $\ell\in\ZZ$ and $\xi$ is an independent parameter.
The crossing parameter is $\lam$ and the spectral parameter is $u$.
The function $h(u)$ is given by
\be
h(u)=\Theta(0)H(u)\Theta(u) \; .
\ee
The face weights satisfy the following Yang-Baxter equation
\be
&&\sum_g\wt Wabgfu\wt {W}fgdev\wt {W}gbcd{v\!-\!u} \no \\
&&\h=\sum_g\wt {W}fage{v\!-\!u}\wt {W}abcgv\wt Wgcdeu \label{STR}
\ee
for any integers $a,b,c,d,e,f.\;$ Therefore the SOS model is an
integrable system. Suppose that \mbox{\boldmath $l$} and
\mbox{\boldmath $m$} are allowed spin  configurations of two consecutive
rows of an $N$ ($even$)  column lattice with periodic boundary
conditions $l_{N+1}=l_1$, $m_{N+1}=m_1$. The elements of the row-to-row
transfer matrix $\T$  of the SOS model are defined by
\begin{eqnarray}
\langle\mbox{\boldmath $l$}|{\bf T}(u)|\mbox{\boldmath $m$}\rangle =
\prod_{j=1}^N W(l_j,l_{j+1},m_{j+1},m_{j}|u)\;.
\ee
We recall the eigenvalues of the transfer matrix $\T$ given in
\cite{Baxter:72} (also see \cite{FaTa:79} for algebraic
Bethe ansatz),
\be
T(u)=e^{\i s\lam}h^N(u+\half\lam){q(u-\lam)\over q(u)}
 +e^{-\i s\lam}h^N(u-\half\lam){q(u+\lam)\over q(u)} \label{eigen}
\ee
where $q(u)$ is defined by
\be
q(u)=\prod_{j=1}^{N/2}h(u-u_j) \; .
\ee
These parameters $u_1,u_2,\cdots,u_{N/2}$ are determined by
the Bethe ansatz equations,
\be
p(u_j)=-1\;, \h j=1,2,\cdots,N/2 \label{BA}
\ee
where the function is given by
\be
p(u):=e^{-2\i s\lam}{h^N(u-\half\lam)q(u+\lam)
    \over h^N(u+\half\lam)q(u-\lam)}\; .
\ee
The ABF RSOS model is specialized by setting
\be
\lam=k\pi/h \and \xi=0\;  \label{lam}
\ee
where $k$ and $h$ are relatively prime integers ($h>k>0$) and
$s=1,2,\cdots,h-1$. With this condition (\ref{lam}) the face weights
still satisfy the Yang-Baxter equation. The row-to-row transfer
matrix $\T(u)$ forms the commuting family
\be
\left[\T(u)\; ,\; \T(v)\;\right]=0\;.
\ee
Therefore the model is integrable. The Bethe ansatz solutions
(\ref{eigen}) and (\ref{BA}) with the restriction (\ref{lam}) are the
eigenvalues and the Bethe ansatz equations
of the transfer matrix of the RSOS model \cite{Baxter:82,BaRe:89}.

\section{Finite-size corrections}\label{calculation}\setcounter{equation}{0}

We consider the corresponding critical ABF RSOS model, which can be
obtained by taking the zero elliptic nome $p=0$.
The elliptic function $h(u)$ reduces to the trigonometric function
\be
h(u)=\sin(u) \label{rep}
\ee
if $p\to 0$.
The eigenvalues (\ref{eigen}) and the Bethe ansatz equations (\ref{BA})
are still correct for the critical RSOS model if the function
$h(u)$ is replaced with (\ref{rep}).

Let us introduce the new spectral variable $v=iu$. It is very helpful to
notice that the eigenvalue spectra (\ref{eigen}) and the Bethe ansatz
equations (\ref{BA}) are the same as those of the transfer matrix of the
six vertex model with a twisted boundary condition \cite{KBP:91}. They
therefore can be treated similarly. The functions have to be restricted
in some analyticity domain since all functions are $\i\pi$-periodic. It
has been shown in \cite{KBP:91} that the following functions
are {\it analytic} and {\it non-zero} ({\sc anz})
\be
 h(v) & \mbox{ {\sc anz} \h in}\;&  0<\im(v)<\pi           \no \\
 q(v) & \mbox{ {\sc anz} \h in}\;&  -\pi<\im(v)<0          \no \\
 T(v) & \mbox{ {\sc anz} \h in}\;&  -\lam/2\le \im(v)\le\lam/2\;,
\label{reg}\ee
and, respectively, the functions $q$ and $p$ satisfy
\be
\ol{q}(v)=q(\ol{v}) \and \ol{p}(v)=1/p({v}) \; .\label{qp}
\ee
Because of $\i\pi$-periodic functions of face weights
we can take $k<h/2$. Note that
(\ref{reg}) has restricted the model to stay on the  critical
 line of regime \reg.

\subsection{Nonlinear integral equation}

Following \cite{KBP:91}, let us introduce new functions
\be
\a (x)&:=&1/p(x-\i\lam/2)\;=\;
    \left[\tanh{\pi x\over 2\lam}\right]^N a(x)\no \\
 &&\U (x)\;:=\;1+\a (x) \label{def-a}
\ee
The variable $x$ may be regarded as real.\footnote{Sometimes it is
convenient to work with values of $x$ in the upper half plane close to
the real axis for avoiding singularities which might otherwise occur.}
The method presented in \cite{KBP:91} is to derive a set of relations
about functions $a$ and $q$ and these relations lead to a nonlinear
integral equation, in turn, the nonlinear integral equation ensures that
the finite-size corrections to the eigenvalue spectra of
the transfer matrix can be solved through dilogarithmic functions.

The involved functions are  {\sc anz} in the  strips (\ref{reg})
and are exponentials in asymptotic behaviour. The second logarithmic
derivatives of the functions can be Fourier transformed,
\be
f(k)&=&{1\over 2\pi}\int_{-\infty}^{\infty}
                   [\ln f(x)]^{''}e^{-\i k x}dx\no \\
\h[\ln f(x)]^{''}&=&\int_{-\infty}^{\infty}\;f(k)\;e^{\i k x}\;dk
\ee
where the integration path in $x$-plane has to lie in the analyticity
strip and the real part of the variable of integration goes from
$-\infty$ to $\infty$. By Cauchy's theorem all other details of the path
are irrelevant for $f(k)$.

We now derive a set of relations about functions $a$ and $q$. Applying
the Fourier transform to the definition (\ref{def-a}) of $a(x)$
\be
a(x)=e^{2\i s\lam}\left[\coth{\pi x\over 2\lam}\right]^N{h^N(x)q(x-3\i\lam/2)
    \over h^N(x-\i\lam+\i\pi)q(x+\i\lam/2-\i\pi)}\;
\ee
where all arguments of the functions $q$ and $h$ have been reduced to
the analyticity strips (\ref{reg}) because of the $\pi\i$-periodicity,
then yields
\be
a(k)&=& -{Nk\over 1+e^{-\lam k}}+{Nk(1-e^{(\lam-\pi)k})\over 1-e^{-\pi k}}
+\left(e^{\talf\lam k}-e^{(\pi-\half\lam)k}\right) q(k)\;. \label{a-q}
\ee
To solve $a$ and $q$ we need another relation, which is introduced by
an auxiliary function
\be
h_{\mbox{\scr a}}(v):={1+p(v)\over q(v)}\;.
\ee
It is {\sc anz} in the strip $-\lam/2<\im (v)\le\lam/2$. To apply
Cauchy's theorem to the Fourier transform of the second logarithmic
derivative of $h_a$ we rewrite $h_{\mbox{\scr a}}(v)$ in the following
two different forms such that the arguments of $q$ stands in the
analyticity strip (\ref{reg})
\be
h_{\mbox{\scr a}}(x-\i\lam/2)&=&\left[\coth{\pi x\over 2\lam}\right]^N
{\U(x)\over
       q(x-\i\lam/2)\;a(x)} \no \\
h_{\mbox{\scr a}}(x+\i\lam/2)&=&{\ol{\U}(x)\over q(x+\i\lam/2-\i\pi)} \;.
\label{h}\ee
Applying the Fourier transform to (\ref{h}), it follows that
\be
e^{\lam k/2}\;h_{\mbox{\scr a}}(k)&=&-{Nk\over 1+e^{-\lam k}}
        -e^{\lam k/2}\;q(k)+\U(k) -a(k)\no\\
e^{-\lam k/2}\;h_{\mbox{\scr a}}(k)
      &=&\ol{\U}(k)-e^{(\pi-\lam/2) k}\;q(k) \;.\no
\ee
Then they are equated yielding
\be
q(k)\left(e^{(\pi+\lam/2) k}-e^{\lam k/2}\right)&=&
  {Nk\over 1+e^{-\lam k}}+a(k)+e^{\lam k}\ol{\U}(k)-{\U}(k)\;.\label{q-a}
\ee
The equations (\ref{a-q}) and (\ref{q-a}) together determine the functions
$a(k)$ and $q(k)$,
\be
a(k)&=&{\sinh(\half\pi k-\lam k)\over
    2\cosh(\half\lam k)\sinh\half(\pi k-\lam k)}
 \left(\U(k)-e^{(\lam-\epsilon)k}\ol{\U}(k)\right) \no \\
q(k)&=&{Nke^{-\pi k/2}\over 4\sinh(\half\pi k)\cosh(\half\lam k)} \no \\
    &&-{e^{-(\pi+\lam)k/2}\over 4\cosh(\half\lam k)\sinh\half(\pi k-\lam k)}
     \left(\U(k)-e^{\lam k}\ol{\U}(k)\right) \label{q(k)}
\ee
where an infinitesimal positive $\epsilon$ has been introduced
for the imaginary part of the argument of $x$.
Transforming back to the variable $x$
\be
\;[\ln a(x)]^{''}&=&\int_{-\infty}^{\infty}\left(K(y)[\ln\U]^{''}(x-y)-
      K(y+\i\epsilon-\i\lam)[\ln\ol{\U}]^{''}(x-y)\right)dy
\label{aa}\ee
where the kernel function
\be
K(x):={1\over 2\pi}\int_{-\infty}^{\infty}{\sinh(\half\pi-\lam)k\over
     2\cosh(\half\lam k)\sinh\half(\pi-\lam)k}\;e^{\i kx}\;dk\;
\ee
satisfies
\be
\ol{K}(x)=K(-\ol{x})\;,\h K(x)=K(-x)\;.
\ee
The equation (\ref{aa}) is derived based on the essential {\sc anz}
property of  the Bethe ansatz solution (\ref{eigen}). Low-lying
excitations have  the same bulk behavior as the ground state.
The only difference has been shown in \cite{KlPe:91} to lie in the fact
that the eigenvalue functions now possess a finite number of zeros in the
analyticity strip, which were free of zeros in the ground state. However,
it is always possible to take an {\sc anz} area in the analyticity strip
where Cauchy's theorem can be applied \cite{KlPe:92}. Therefore the equation
(\ref{aa}) still works for the excited states if we change the
integration path in the {\sc anz} area. Integrating (\ref{aa})
twice we obtain a nonlinear integral equation
\be
\ln a(x)&=& \int_{-\infty}^{\infty}\left(K(y)\ln\U(x-y)-
     K(y+\i\epsilon-\i\lam)\ln\ol{\U}(x-y)\right)dy \no\\
 &&\h\h +C+Dx \label{cd}
\ee
where the integral constant $D=0$ because all terms remain
finite for $x\to\infty$ and another integral constant $C$
heavily dependent on the branch choice of  $\ln a(x)$,
\be
C&=&\ln a(\infty)-\int_{-\infty}^{\infty}K(y)dy\left(\ln\U(\infty)-
          \ln\ol{\U}(\infty)\right) \no \\
&=&\ln(\omega^2e^{2\i s\lam})-{\half\pi-\lam\over \pi-\lam}
   \left(\ln(1+\omega^2e^{2\i s\lam})-
    \ln(1+\omega^{-2}e^{-2\i s\lam})\right) \no \\
&=&{\i\pi\phi\over \pi-\lam}
\ee
where the phase factor $\phi$ has been introduced by
\be
\phi=s\lam-\i\ln\omega\;, \h \omega^2=1\; .
\ee
Here we have taken a more general choice of branches for $\ln a(x)$,
or $\ln a(\infty)=2\ln(\omega e^{\i s\lam})$. The case  $\omega=1$
has been studied in \cite{KBP:91}, which corresponds to the
ground state.  For the excited states we have chosen the other
branches with $\omega\not=1$ or take
\be
\omega=e^{\i (t-s)\pi}
\ee
with the integers $s,t$.  From the definition (\ref{def-a}) it follows that
$a$ goes to $a\;e^{-4\i s\lam}$ under the change $s\to h-s$.
So the same symmetry should be imposed on the equation (\ref{cd}),
or the phase factor $\phi$ must go to $-\phi$ under this change.
It follows that
the phase factor $\phi$  will go to  $-\phi$  if  changing
$s\to h-s$ and $t\to h-k-t$. Similar to the exponent $s$, suppose
that $t$ is positive. According to $s=1,2,\cdots,h-1$
we therefore take   $t=1,2,\cdots,h-k-1$.
Recalling the definition (\ref{def-a}) we arrived at the nonlinear
integral equation for $\a$
\be
\ln\a(x)&=&N\ln\tanh{\pi x\over 2\lam}+{\i\pi\phi\over \pi-\lam}\no\\
&& +\int_{-\infty}^{\infty}\left(K(y)\ln\U(x-y)-
     K(y+\i\epsilon-\i\lam)\ln\ol{\U}(x-y)\right)dy \;.\label{a}
\ee
This equation is exact for all finite system size and for both the
ground state and the excited states.

\subsection{Scaling limits}
To obtain the finite-size corrections to the eigenvalue spectra of the
transfer matrix we observe the following scaling behaviour
\be
\lim_{N\to\infty}\left(\tanh\left[{\pi\over 2\lam}\(\pm{\lam\over\pi}
  (x+\ln N)\)\right]\right)^N=\exp\left(-2e^{-x}\right)\;
\ee
in thermodynamic limit $N\to\infty$. The function $\a$ scale similarly,
\be
&a_\pm(x):=\disp{\lim_{N\to\infty}}\a\(\pm{\lam\over\pi}(x+\ln N)\)\;,
  &\la_\pm(x):=\ln a_\pm(x)  \\
&A_\pm(x):=\disp{\lim_{N\to\infty}}\U\(\pm{\lam\over\pi}(x+\ln N)\)
 =1+a_\pm(x)\;,  & \lA_\pm(x):=\ln A_\pm(x)\;.  \label{scal}
\ee
In the scaling limit regimes the nonlinear integral equation (\ref{a})
becomes
\be
\la_\pm(x)&=&-2e^{-x}+\int_{-\infty}^{\infty}K_1(x-y)\lA_\pm(y)du \no \\
&& -\int_{-\infty}^{\infty}K_2(x-y)\ol{\lA}_\pm(y)du +
    {\i\pi\phi\over \pi-\lam}       \no\\
\ol{\la}_\pm(x)&=&-2e^{-x}+\int_{-\infty}^{\infty}\ol{K}_1(x-y)
         \ol{\lA}_\pm(y)du      \no \\
&&-\int_{-\infty}^{\infty}\ol{K}_2(x-y){\lA}_\pm(y)du -
    {\i\pi\phi\over \pi-\lam} \label{aa-1}
\ee
where $K_{1,2}(x)$ are defined by
\be
K_{1}(x)&:=&{\lam\over \pi}K\left({\lam\over \pi}x\right) \no\\
K_{2}(x)&:=&{\lam\over \pi}K\left({\lam\over
\pi}x\pm\i(\epsilon-\lam)\right)\;.
\ee

Let us now turn to the eigenvalues $T$ given by (\ref{eigen}).
Its finite-size
corrections can be derived from
\be
T(x-\i\lam/2)=h^N(x-\i\lam){q(x+\i\lam/2-\i\pi)\over q(x-\i\lam/2)}
   {\U(x)}e^{-s\i\lam}\;.
\ee
Applying the Fourier transform to the ratio of the $q$-functions
and taking (\ref{q(k)}) into account
we have
\be
\ln{q(x+\i\lam/2-\i\pi)\over q(x-\i\lam/2)}&=&-N\int_{-\infty}^{\infty}
     {\sinh\half(\pi-\lam)k\over 2k\sinh\half(\pi k)\cosh\half(\lam k)}
      e^{\i kx}dy     \no \\
&&\hs{-2.5}+{\i\over 2\lam}\int_{-\infty}^{\infty}{\ln\U(x-y)\over
     \sinh{\pi\over\lam}(y-\i\epsilon)}dy+{\i\over 2\lam}
    \int_{-\infty}^{\infty}{\ln\ol{\U}(x-y)\over
     \sinh{\pi\over\lam}(y+\i\epsilon)}dy +f_c
\ee
where $f_c$ is an integration constant.
Therefore the finite-size corrections to the eigenvalue can be
expressed as
\be
\ln T(x-\i\lam/2)&=&f_c+N\ln h(x-\i\lam)-N\int_{-\infty}^{\infty}
    {\sinh\half(\pi-\lam)k\sinh(xk)\over
     2k\sinh\half(\pi k)\cosh\half(\lam k)}dk \no \\
&& +  {\i\over \lam}\int_{-\infty}^{\infty}{\re\ln\U(y)\over
     \sin{\pi\over\lam}(x-y+\i\epsilon)}dy+{ o\!}\({1\over N}\)\;.
\ee
The scaling limit of the corrections can be done by splitting the
integral into two parts, then inserting the variable of integration
$y$ by $\pm{\lam\over\pi}(y+\ln N)$ and using the scaling functions
(\ref{scal}), we obtain
\be
&&\hs{-0.3} \ln T(x-\i\lam/2)\no \\
&&=-Nf(x-\i\lam/2)-{2\i\over\pi N}e^{{\pi\over\lam}x}
  \int_{-\infty}^{\infty}\re\lA_+(y)e^{-y}dy \no \\
&&\h \h +{2\i\over\pi N}e^{-{\pi\over\lam}x} \int_{-\infty}^{\infty}
      \re\lA_-(y)e^{-y}dy+{ o\!}\({1\over N}\)\no\\
&&=-Nf(x-\i\lam/2)-{\pi\i\over 6 N}\sinh{\pi x\over\lam}\(
   {24\over\pi^2}\int_{-\infty}^{\infty}\re\lA_{\pm}(y)e^{-y}dy\)
       +{ o\!}\({1\over N}\)\; \label{finite}
\ee
where the bulk behavior is entirely expressed by the first term
and second term is the finite-size corrections. The integration
constant $f_c$ is chosen so that $f(x-\i\lam/2)$ is exactly the
bulk energy, which can be derived from the inversion relation of
the face weights \cite{Baxter:82,FoBa:85}. Here we are only
interested in the finite-size correction terms which include
the conformal spectra.

\subsection{Conformal spectra}
The conformal spectra can be extracted from the
finite-size corrections of the transfer matrix.  The integral
in the finite-size correction term in (\ref{finite}) can be
calculated by considering the expression
\be
&&\int_{-\infty}^{\infty}\left([\la_\pm(x)]^{'}\lA_\pm(x)-
                  \la_\pm(x)[\lA_\pm(x)]'\right)dx \no \\
&&\h\h    +\int_{-\infty}^{\infty}\left([\ol{\la}_\pm(x)]'\ol{\lA}_\pm(x)-
                  \ol{\la}_\pm(x)[\ol{\lA}_\pm(x)]'\right)dx \no \\
&&=2\int_{-\infty}^{\infty}e^{-x}
       \left({\lA}_\pm(x)+[{\lA}_\pm(x)]^{'}\right)dx\no \\
&&\h + 2\int_{-\infty}^{\infty}e^{-x}\left(\ol{\lA}_\pm(x)
                          +[\ol{\lA}_\pm(x)]'\right)dx\no \\
&&\h -{\pi\i\phi\over\pi-\lam}
      \int_{-\infty}^{\infty}
       \left([{\lA}_\pm(x)]'-[\ol{\lA}_\pm(x)]'\right)dx\;.
\ee
The right hand side is derived by using the nonlinear integral
equation (\ref{aa-1}). The left hand side can be calculated after
changing the variable $x$ to $a$ and $\ol a$ and
using the dilogarithmic function
\be
L(x)=\int^x_0\left({\ln(1+y)\over y}-{\ln y\over 1+y}\right]dy\;.
\ee
Then we are able to derive
\be
&&{24\over\pi^2}\int_{-\infty}^{\infty}e^{-x}\re\lA_\pm(x)dx \no \\
&&={3\over\pi^2}\left(L\left(\omega^2e^{2\i s\lam}\right)+
     L\left(\omega^{-2}e^{-2\i s\lam}\right)
    -{2\pi\phi^2\over\pi-\lam}\right)
\ee
where the asymptotics of $a_\pm(\infty)=\left(\omega e^{\i s\lam}
\right)^2$, $\ol{a}_\pm(\infty)=\left(\omega e^{\i s\lam}\right)^{-2}$
and $a_\pm(-\infty)=\ol{a}_\pm(-\infty)=0$ have been read
off from (\ref{aa-1}). Finally using the well known identity
\be
L(z)+L(1/z)={\pi^2\over 3}
\ee
the finite-size corrections in (\ref{finite}) are given by the explicit
expression
\be
\ln T(x-\i\lam/2)=-Nf(x-\i\lam/2)-
 {\pi\i\over 6N}\(c-24\Delta\)\sinh{\pi x\over\lam}
  +{ o\!}\({1\over N}\)
\ee
or changing the variable $x$ to $v=\i u=x-\i\lam/2$
\be
\ln T(v)=-Nf(v)+
 {\pi\i\over 6N}\(c-24\Delta\)\cosh{\pi v\over\lam}
  +{ o\!}\({1\over N}\)\; ,
\ee
where the central charge is
\be
c=1-{6\lam^2\over \pi(\pi-\lam)} \label{c}
\ee
and the conformal weights are
\be
\Delta={\phi^2-\lam^2\over 4\pi(\pi-\lam)} \;.  \label{delta}
\ee

For the ground state, $s=t=1$ yields $\Delta=0$. The choice of
$1<s\le h-1$ and $1<t\le h-k-1$ gives the excited states. Remarkably,
inserting
$\lam$ given by (\ref{lam}) into the conformal spectra we have the
central charges and the conformal weights of the primary fields for
Virasoro minimal models
\be
&&c=1-{6k^2\over h(h-k)} \h\and\h\Delta={[ht-(h-k)s]^2-k^2\over 4h(h-k)}  \\
&& \h  k<h; \h s=1,2,\cdots,h-1;\; t=1,2,\cdots,h-k-1 \no
\ee
for $k\ge 1$. The unitary minimal models are given by taking $k=1$.

\section{Discussion}\setcounter{equation}{0}

In this paper we have obtained the conformal spectra of the
non-unitary minimal conformal field theories
from the finite-size corrections to the eigenvalue spectra of
the transfer matrix of the
critical ABF model on the regime  \reg critical line
with the crossing parameter (\ref{lam}).
The method given in \cite{KBP:91} is only for calculating the central
charges for the six-vertex model with a twisted boundary condition.
In this paper it has been generalized to calculate both the central
charges and the conformal weights. Other methods, for example, the
thermodynamic Bethe ansatz (TBA) analysis (see
\cite{BaRe:89}, \cite{YY:69}-\cite{Kuniba:93}),
%,Tak:71,TaSu:72,Zam:91,KlMe:90,KlMe:91,Martins:91,
exist for calculating the conformal spectra. The TBA relies heavily on
the string hypothesis, while our method crucially depends on
the {\sc anz} property instead. However, it is an interesting problem to
generalize the TBA method for calculating the
conformal weights of the ABF SOS model.

There is another method for calculating the finite-size corrections
of transfer matrices. This has been shown by solving the fusion
hierarchies of the ABF model. Unfortunately it is only for $k=1$
\cite{KlPe:92} (also see \cite{ZhPe:95}). It has not yet known how to
find finite-size corrections
of the transfer matrix of the ABF model for $k>1$.

\bigskip
\noindent
{\bf Note Added}

\noindent
After this work was submitted I was informed by Murray Batchelor about
reference \cite{KWZ:93}, where the authors generalize the method of
\cite{KBP:91} to
 calculate the conformal spectrum of the six-vertex model with twisted
 boundary conditions. However, the finite-size corrections to
the transfer matrix of the ABF SOS model with the more general crossing
parameter (1.12), which is an important class of integrable lattice models
corresponding to realizations of non-unitarity minimal conformal field
theories,
was not considered there.
I am grateful to Murray Batchelor for drawing my attention to \cite{KWZ:93}.

\section*{Acknowledgements} This research has been supported by the
Australian Research Council. The author also thanks P. A.
Pearce and Ole Warnaar for discussions.

\end{document}